\magnification=1200
\settabs 18 \columns

\baselineskip=12 pt
\topinsert \vskip 1.25 in
\endinsert

\def\sqr#1#2{{\vcenter{\vbox{\hrule height.#2pt
 \hbox{\vrule width.#2pt height#1pt \kern#1pt
 \vrule width.#2pt} \hrule height.#2pt}}}}

\def\operp{\hbox{${\kern+.25em{\bigcirc}
\kern-.85em\bot\kern+.85em\kern-.25em}$}}

\def\lsim{\;\raise0.3ex\hbox{$<$\kern-0.75em\raise-1.1ex\hbox{$\sim$}}\;}
\def\gsim{\;\raise0.3ex\hbox{$>$\kern-0.75em\raise-1.1ex\hbox{$\sim$}}\;}
\def\no{\noindent}

\def\ce{\centerline}
\def\ve{\vfill\eject}
\def\rdots{\mathinner{\mkern1mu\raise1pt\vbox{\kern7pt\hbox{.}}\mkern2mu
 \raise4pt\hbox{.}\mkern2mu\raise7pt\hbox{.}\mkern1mu}}

\def\e e{$e^+ e^-$ }



\rightline{UCLA/00/TEP/24}
\rightline{August 2000}
\vskip1.0cm

\ce{{\bf TCP in $q$-Lorentz Theories}}
\vskip.5cm
\ce{\it R. J. Finkelstein}
\vskip.3cm
\ce{Department of Physics and Astronomy}
\ce{University of California, Los Angeles, CA  90095-1547}
\vskip1.0cm

\no {\bf Abstract.}  The connection between spin and statistics implied by
the continuous Lorentz group together with strong reflection (TCP) is shown
to hold also for the $q$-Lorentz group.
\ve

\line{{\bf 1. Introduction.} \hfil}

The relation of Lorentz invariance to Bose-Einstein and Fermi-Dirac
statisticis was discussed by Pauli in several investigations leading to a
formulation bsed on the principle of invariance under strong reflection
(TCP invariance)$^1$ as formulated by Schwinger.$^2$  There is nothing in these studies, however, to
suggest how the commutators of bose fields or the anticommutators of fermion
fields would be altered if the Lorentz group were deformed and replaced by
another group.  This question is now of some interest in connection with
speculations about new statistics associated with $q$-groups.  We shall here
examine the problem for the $q$-Lorentz group.
\vskip.5cm

\line{{\bf 2. The $q$-Lorentz group and the possibility of $q$-Commutators.} \hfil}

The two-dimensional representation of the Lorentz group $(L)$ satisfies
$$
L^t\epsilon L = L\epsilon L^t = \epsilon \eqno(2.1)
$$
\no where $L^t$ is the transposed $L$ and $\epsilon$ is the Levi-Civita
symbol
$$
\epsilon_{ij} = \left(\matrix{0 & 1\cr -1 & 0 \cr} \right)~.
\eqno(2.2)
$$

We define the two-dimensioinal representation of the $q$-Lorentz group
$(L_q)$ by a similar relation:
$$
L_q^t\epsilon_q L_q = L_q\epsilon_q L_q^t = \epsilon_q \eqno(2.3)
$$
\no where
$$
\epsilon_q = \left(\matrix{0 & q_1^{1/2}\cr -q^{1/2} & 0\cr}\right)
\qquad q_1 = q^{-1} \eqno(2.4)
$$
\no The two groups then agree in the limit $q=1$.

To illustrate our question let us consider a vector-spinor field with the
conventional Lorentz invariant interaction
$$
\bar\psi A\!\!\!/\psi
$$
\no and let us pursue conventional theory aside from the introduction of
the following time ordered products:
$$
\eqalign{T_q(\psi(x)\psi(x^\prime)) &=\psi(x) \psi(x^\prime) \qquad
t>t^\prime \cr
&= q^\prime\psi(x^\prime)\psi(x) \quad t<t^\prime~~{\rm vector} \cr
&= q^{\prime\prime}\psi(x^\prime)\psi(x) \quad
t<t^\prime~~{\rm spinor} \cr} 
$$

Then the $q$-time ordered $S$-matrix is
$$
S^{(q)} = T_q(e^{i\int{\cal{L}}(x) d^4x}) 
$$
\no where $(q)=(q^\prime,q^{\prime\prime})$, and the Wick expansion leads to
Feynman rules with the following $q$-propagators$^3$
$$
\eqalign{D^{q^\prime}_{\mu\lambda}(x) &=
\biggl(g_{\mu\lambda}-{\partial_\mu\partial_\lambda\over m^2}\biggr)
\biggl({1\over 2\pi}\biggr)^4\biggl({1+q^\prime\over 2}\biggr) 
\int e^{-ikx}{1\over k^2-m^2}\biggl[1+{1-q^\prime\over 1+q^\prime}
{k_o\over\omega}\biggr] d^4k  \cr
S_{\alpha\beta}^{q^{\prime\prime}}(x) &=
(\partial\!\!\!/+m)_{\alpha\beta}\biggl({1\over 2\pi}\biggr)^4
\biggl({1-q^{\prime\prime}\over 2}\biggr) \int e^{-ikx}
{1\over k^2-m^2}\biggl[1+{1+q^{\prime\prime}\over 1-q^{\prime\prime}}
{k_o\over\omega}\biggr] d^4k~.  \cr}
$$
\no The difference beween these two propagators comes from  $(q^\prime,q^{\prime\prime})$ and the sum over polarization and antiparticle
states.  The final theory may be tested for Lorentz invariance by calculating
particle-particle scattering which depends on the photon propagator, and
particle-antiparticle annihilation, which depends on the spinor propagator.

In both cases the result is frame dependent, i.e., Lorentz invariance is
broken unless $q^\prime=1$ for the vector and $q^{\prime\prime}=-1$ for the
spinor propagator.  This result is then a special case of the Pauli theorem
that Lorentz invariance requires commutation and anticommutation rules for
integer and half-integer spin respectively.

Our question is now: if the $L$ group is replaced by $L_{q^\prime}$, are
$q^\prime$ and $q^{\prime\prime}$ still constrained to be +1 and -1 or will
one find two new functions: $q^\prime=q^\prime(q)$ and
$q^{\prime\prime}=q^{\prime\prime}(q)$?
\vskip.5cm

\line{{\bf 3. Strong Reflection.} \hfil}

In the van der Waerden representation of the Lorentz group one utilizes
two spinor representations related by complex conjugation, permitting the
general tensor to be written as
$$
u(n,m) = u_{k_1}\ldots u_{k_n}~v_{\dot\ell_1}\ldots v_{\dot\ell_m} \eqno(3.1)
$$
\no where the dotted index is the notation for complex conjugation.  The
tensor (3.1) transforms like the product of $n$ spinors and $m$ complex
conjugate spinors.  A field characterized by $u_{nm}(x)$ is called fermionic
if $n+m$ = odd and bosonic if $n+m$ = even.

The Pauli formulation of strong reflections makes use of the following
transformation laws for a general tensor field $u(n,m)$
$$
u^\prime(n,m) = i(-1)^n u(n,m) = -i(-)^m u(n,m) \eqno(3.2)
$$
\no for fermionic fields ($n+m$ = odd) and
$$
u^\prime(n,m) = (-)^n u(n,m) = (-)^m u(n,m) \eqno(3.3)
$$
\no for bosonic fields ($n+m$ = even).

The transformations (3.2) and (3.3) preserve the reality condition
$$
u(n,m)^* = v(m,n)
$$
\no where * means complex conjugation.

To complete the prescription for strong reflection, one must reflect all
coordinates and reverse the order of all fields appearing in a product of
several fields.  Strong reflection will thus preserve a chronological order
of factors.

If the fields appearing in the product are all fermionic, then by applying
(3.2) $N$ times we have
$$
\eqalign{\bigl((u_{n_1m_1}(x_1)\ldots u_{n_Nm_N}(x_N)\bigr)^\prime &= ~i^N(-)^{n_1+\ldots+n_N}\bigl(u_{n_1m_1}(x_1)\ldots u_{n_Nm_N}(x_N)\bigr) \cr
\noalign{\vskip3pt}
\noalign{\hbox{After reversing all coordinates and order of all fields, one 
has}}
\noalign{\vskip3pt} 
&= fi^N(-)^{n_1+\ldots + n_N}
\bigl(u_{n_Nm_N}(-x_N)\ldots u_{n_1m_1}(-x_1)\bigr) \cr} \eqno(3.5)
$$
\no where $f$ is a complex number introduced by reversing the order of the
fields in the product and has still to be fixed.

If the individual fields $u_{n_km_k}(x)$ are fermionic, then the product
fields are also either fermionic or bosonic depending on whether $N$ is odd
or even, since we may write
$$
n = \sum^N_1 n_k \qquad m = \sum^N_1 m_k
$$
\no and
$$ 
\eqalign{n+m = \sum^N_1 (n_k+m_k) =
\sum^N_1~\hbox{(odd)$_k$}~&= ~\hbox{(odd)$_k$ if}~N~ \hbox{is odd} \cr
&=~\hbox{(even) if}~N~ \hbox{is even}~. \cr}
$$

In (3.5) set
$$
u_{nm} = u_{n_1m_1} \ldots u_{n_Nm_N}~.
$$
\no Then $u_{nm}$ is either a bosonic or a fermionic field and must
satisfy (3.3) or (3.2).

For agreement between (3.5) and (3.2) we require
$$
f = i^{1-N} \qquad N~\hbox{odd}  \eqno(3.6)
$$
\no and between (3.5) and (3.3)
$$
f = i^{-N} \qquad N~\hbox{even} \eqno(3.7)
$$
\no If $N$ = (odd,even) set $N=(2p+1,2p)$ then $f=(i^{-2p},i^{-2p})$ and
note
$$
\eqalign{(-1)^{{N\over 2}(N-1)} &= i^{2p(2p-1)} = i^{-2p} 
\qquad~~~~~ N~\hbox{even} \cr
&= i^{(2p+1)(2p)} = i^{2p} = i^{-2p}~~N~\hbox{odd} \cr} \eqno(3.8)
$$
\no Hence in both cases
$$
f = (-1)^{{N\over 2}(N-1)}~. \eqno(3.9)
$$

If one requires that the order $u_{n_Nm_N}(x_N)\ldots u_{n_1m_1}(x_1)$ be
obtained from \break $\bigl((u_{n_1m_1}(x_1)\ldots u_{n_Nm_N}(x_N)\bigr)$ as a
product of transpositions between neighboring factors each of which
introduces the same factor $\eta$ then
$$
f = \eta^{{N\over 2}(N-1)} \eqno(3.10)
$$
\no as well as (3.9).  For $N=2~\eta=-1$.  Since $\eta$ has the same value
for all transpositions, $\eta = -1$ for all $N$.

If there are boson fields in the product there are similar simpler remarks.

In this way strong reflection invariance implies the usual commutation and
anticommutation rules between fields.
\ve

\line{{\bf 4. $q$-Tensors.} \hfil}

The general $q$-tensor is now defined as a product of spinors that copies
the product (3.1), namely:
$$
u_{k_1\ldots k_n}^{\dot\ell_1\ldots\dot\ell_m} = u(k_1) \ldots u(k_n)
v(\dot\ell_1)\ldots v(\dot\ell_n) \eqno(4.1)
$$
\no where $u$ and $v$ are fundamental representations of $L_q$ and
$\dot L_q$ defined as follows:
$$
\eqalignno{&\epsilon_q = L_q^t\epsilon_q L_q = L_q\epsilon_q L_q^t  &
(4.2) \cr 
\noalign{\hbox{and}}
&\dot\epsilon_q = \dot L_q^t\dot\epsilon_q\dot L_q = \dot L_q\dot\epsilon_q\dot L_q^t 
& (4.3) \cr}
$$
\no where the dot again means complex conjugatiion and $\epsilon_q$ is given
by (2.4).  We are interested in the case where $q$ is real.  Then
$$
\dot\epsilon_q = \epsilon_q \eqno(4.4)
$$
$$
\epsilon_q = \dot L_q^t\epsilon_q\dot L_q = \dot L_q\epsilon_q\dot L_q^t
\eqno(4.5)
$$
\no One may carry over the usual notation, i.e. (4.1) may be written
$$
u_{k_1\ldots k_n}^{\dot\ell_1\ldots\dot\ell_m} = u(n,m) \eqno(4.6)
$$
\no where $n+m =$ (even, odd) means (bosonic, fermionic).  One also has
$$
\dot u(n,m) = u(m,n)~. \eqno(4.7)
$$
\no To describe strong reflections we retain (3.2) and (3.3).  These transformations preserve
$$
\dot u(n,m) = v(m,n)~.
$$
\no The argument of the previous section may now be repeated with no changes
if $u(n,m)$ is defined by (4.6).  Our conclusion is that there is no change
in the quantum statistics, as determined by the field commutators and
anticommutators, in passing from $L$ to $L_q$.

\vskip.5cm

\line{{\bf 5. The $q$-Light Cone.$^3$} \hfil}

Since earlier discussions of the connection between spin and statistics
made use of the distinction between causal and non-causal commutators, it is
of interest to see whether the light cone and associated space-like interval
are changed in the $q$-theory.

We summarize a few elements of the $q$-spinor calculus.  The contravariant
$\epsilon$ metric is
$$
\epsilon^{AB}(q) = \epsilon_{AB}(q_1)~. \eqno(5.1)
$$
\no One may define covariant $\sigma$ matrices
$$
(\sigma_q^n)_{B\dot Y} = (1,\vec\sigma)_{B\dot Y} \eqno(5.2)
$$
\no where $\vec\sigma$ abbreviates the usual set of Pauli matrices.  Then
the matrices contravariant to $(\sigma_q^m)_{B\dot Y}$ are
$$
(\bar\sigma_q^m)^{\dot XA} = \epsilon_q^{\dot X\dot Y}\epsilon_q^{AB}
(\sigma_q^m)_{B\dot Y} \eqno(5.3)
$$
\no and the $q$-metric $(\eta)$ is
$$
2\eta^{nm} = (\bar\sigma_q^m)^{\dot XA}(\sigma_q^\eta)_{A\dot X}~. \eqno(5.4)
$$
\no One finds
$$
\eta^{mn} = \left(\matrix{{1\over 2}(q+q_1) & 0 & 0 & {1\over 2}(q-q_1) \cr
0 & -1 & 0 & 0 \cr 0 & 0 & -1 & 0 \cr
-{1\over 2}(q-q_1) & 0 & 0 & -{1\over 2}(q+q_1) \cr} \right) \eqno(5.5)
$$
\no The equation of the light cone is then
$$
\eta_{mn}X^mX^n = \eta_{00}c^2t^2 + \eta_{33}z^2-x^2-y^2 = 0
$$
\no or
$$
{1\over 2}(q+q_1)c^2t^2-{1\over 2}(q+q_1)z^2-x^2-y^2 = 0 \eqno(5.6)
$$
\no or
$$
c^q\tau^2-\zeta^2-x^2-y^2=0 \eqno(5.7)
$$
\no after the rescaling
$$
\eqalign{\zeta &= {1\over 2}(q+q_1)z \cr
\tau &= {1\over 2}(q+q_1)t \cr} \eqno(5.8)
$$
\no The light cone is then simply rescaled in the $q$-theory and so the earlier
discussion about the connection between spin and statistics should hold,
in agreement with the conclusion based on the TCP invariance.

Although $L$ and $L_q$ lead to the same connection between spin and
statistics, there are in principle detectable physical differences between
the two theories coming from the replacement of the charge conjugation
matrix, $C$, in $L$ by the corresponding matrix $(\epsilon_q)$ in 
$L_q$, and justified by
$$
\eqalign{L^tCL &= C \cr L^t\epsilon_q L &= \epsilon_q \cr}
$$
in the two-dimensional representation.

The replacement of $C$ by $\epsilon_q$ leads to a different relation between
particle and antiparticle and in particular it leads to the substitution
of $q$-bilinears for the usual bilinears.  These changes, although
detectable in principle for weak decays, would probably be masked by
radiative corrections.
\ve

\line{{\bf References.} \hfil}
\vskip.3cm

\item{1.} W. Pauli, ``Exclusion Principle, Lorentz Group and Reflection of
Space-Time and Charge" in {\it Niels Bohr and the Development of Physics},
W. Pauli (ed.), Pergamon Press, New York (1955).
\item{2.} J. Schwinger, Phys. Rev. {\bf 82}, 914 (1951).
\item{3.} R. Finkelstein, Int. J. Mod. Phys. A {\bf 11}, 733 (1996).
\item{4.} R. Finkelstein, J. Math. Phys. {\bf 37}, 953 (1996).

\end
\bye